\documentclass[twocolumn,amssymb,amsmath,aps,pra,groupedaddress,nobibnotes,showpacs]{revtex4}
\usepackage{graphicx}

\begin{document}

\title{Two-Photon Interference without Bunching Two Photons}

\author{Yoon-Ho Kim} \email{kimy@ornl.gov}
\affiliation{Center for Engineering Science Advanced Research, Computer
Science \& Mathematics Division\\ Oak Ridge National Laboratory, Oak Ridge,
Tennessee 37831}

\date{\today}

\begin{abstract}
We report an interference experiment in which the two-photon entangled
state interference cannot be pictured in terms of the overlap and bunching of
two individual photons on a beamsplitter. We also demonstrate that two-photon
interference, or photon bunching effect on a beamsplitter, does not occur if
the two-photon Feynman amplitudes are distinguishable, even though individual
photons do overlap on a beamsplitter. Therefore, two-photon interference cannot
be viewed as interference of two individual photons, rather it should be viewed
as two-photon or biphoton interfering with itself. The results may also be
useful for studying decoherence management in entangled two-qubit systems as we
observe near complete restoration of quantum interference after the qubit
pairs, generated by a femtosecond laser pulse, went through certain
birefringent elements.
\end{abstract}


\maketitle

Two-photon quantum interference effects in spontaneous parametric
down-conversion (SPDC) \cite{klyshko} fields have been playing an important
role from the study of fundamental problems of quantum physics \cite{eprb} to
recent demonstrations of quantum cryptography \cite{gisin} due to the
entanglement between the two down-converted photons.

Among many different quantum interference effects in SPDC, the observation of
null (or close to zero) coincidence counts between the detectors placed at the
two output ports of a beamsplitter, when two photons of SPDC are brought back
together on the beamsplitter from the different input ports at the same time,
has attracted a lot of attention over the years. It was first observed by Shih
and Alley \cite{first} and later by Hong, Ou, and Mandel \cite{second}. This
effect, which we refer to as SA/HOM effect, has the following formal
interpretation: The two-photon interference occurs because the two two-photon
amplitudes leading to a coincidence count (both photons are reflected at the
beamsplitter, r-r, or both photons are transmitted at the beamsplitter, t-t)
become indistinguishable, \textit{even in principle}, and cancel each out when
the photons arrive the beamsplitter simultaneously. Due to the destructive
interference (because each photon accumulates $i$ phase shift upon reflection
at the beamsplitter) between r-r and t-t amplitudes, null coincidence counts
are expected \cite{second,kwiat1}.

This formal interpretation is, however, always accompanied by a physical
picture that two individual photons somehow become bunched together at the
beamsplitter when they arrive at the same time. Since now bunched two photons
leave the beamsplitter from the same output port, null coincidence is expected.
Due to this picture, it is indeed quite common for people to think that two
photons must overlap in time at the beamsplitter for these types of two-photon
interference effects to occur \cite{pittman1}. Such a picture, however, gives
too much credit for SA/HOM effect to a simple linear optical beamsplitter since
it implies some types of local nonlinear interactions.

We now ask: Is the overlap of the two down-converted SPDC photons indeed
necessary for SA/HOM effect? Pittman \textit{et al.} first reported an
experiment which dealt with this question \cite{pittman}. In their experiment,
a delay, which is bigger than the individual photons' coherence times,
introduced to one photon before the beamsplitter is compensated by twice the
delay introduced to its twin photon after the beamsplitter (postponed
compensation). They were then able to observe SA/HOM effect even though the two
photons did not overlap at the beamsplitter. However, the laser which pumps the
SPDC process must have coherence time much bigger than the delay introduced
between the photon pairs for Pittman \textit{et al.}'s scheme to work. In fact,
a cw Argon ion laser, which had several orders of magnitude bigger coherence
time than the delay time, was used in their experiment. Since it is known that
the entangled photon pair of SPDC collectively has the properties of the pump
photon, it may be said that the SPDC photons do overlap at the beamsplitter
within the coherence time of the pump photon in Pittman \textit{et al.}'s
scheme. Thus, Pittman \textit{et al.}'s experiment does not provide us with a
clear answer to the question.

In this paper, we wish to report an experiment which conclusively demonstrates
that the `photons overlapping and bunching at the beamsplitter' picture is not
a valid explanation of general SA/HOM effect (whether `photons' refer to the
pump photons or the SPDC photons). In this experiment, the two
photon-wavepackets not only never overlap at the beamsplitter but also the
arrival time difference between the photon pair at the beamsplitter is much
bigger than the coherence time of the pump photon (pulse). Therefore the
`photon bunching' picture is simply not applicable to this scheme. We also
present an experiment in which the SPDC photons do overlap at the beamsplitter,
but SA/HOM interference does not (and cannot) occur. The quantum mechanical
picture based on in(distinguishability) of `two-photon amplitudes', however,
correctly predicts the presence(absence) of the interference.

\begin{figure}[tbp]
\includegraphics[width=3.2in]{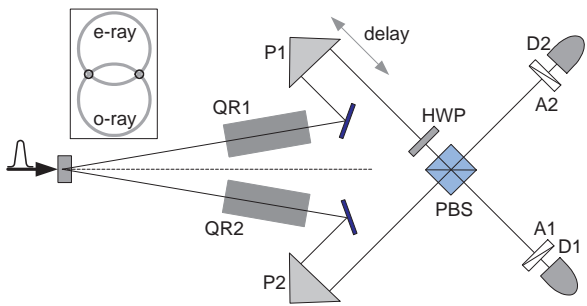}
\caption{\label{fig:exp}Schematic of the experiment. QR1 and QR2 are 20 mm long
quartz rods, HWP is a $\lambda/2$ plate oriented at $45^\circ$. }
\end{figure}

The basic idea of the experiment can be seen in Fig.~\ref{fig:exp}. The photon
pair is generated from a 3 mm thick type-II BBO crystal, with its optic axis
oriented vertically, pumped by an ultrafast laser pulse with coherence time of
approximately 120 fsec. The pump pulse, vertically polarized, has the central
wavelength of 390 nm and the wavelengths of the SPDC photons are centered at
780 nm. As in Ref.~\cite{kwiat}, we consider the intersections of the cones
made by the e- and o-rays exiting the BBO crystal. In each of these two
directions, a photon of either polarization (horizontal or vertical) may be
found, with the orthogonal polarization found in the conjugate photon (i.e.,
individual photons are unpolarized) \cite{note}.

Each photon then passes through a 20 mm long quartz rod (QR1 and QR2), which
generates a relative group delay between the two photons, depending on the
polarization of the photon and the orientation of the optic axis of the quartz
rod. The polarization of one of the photons is then flipped by a $45^\circ$
oriented half-wave plate (HWP). The interferometer is completed by a polarizing
beamsplitter (PBS) and the delay between the two arms is introduced by moving
one of the two trombone prisms (P1 and P2). Photon pairs are then detected by
two single-photon counting modules (D1 and D2) after passing through polarizers
(A1 and A2). In front of each detectors, a 20 nm FWHM interference filter is
introduced to reduce background noise. The outputs from the two detectors were
fed to a time-to-amplitude converter (TAC) and the TAC output was analyzed by a
multi-channel analyzer with a coincidence window set to 3 nsec.

\begin{figure}[tbp]
\includegraphics[width=3.2in]{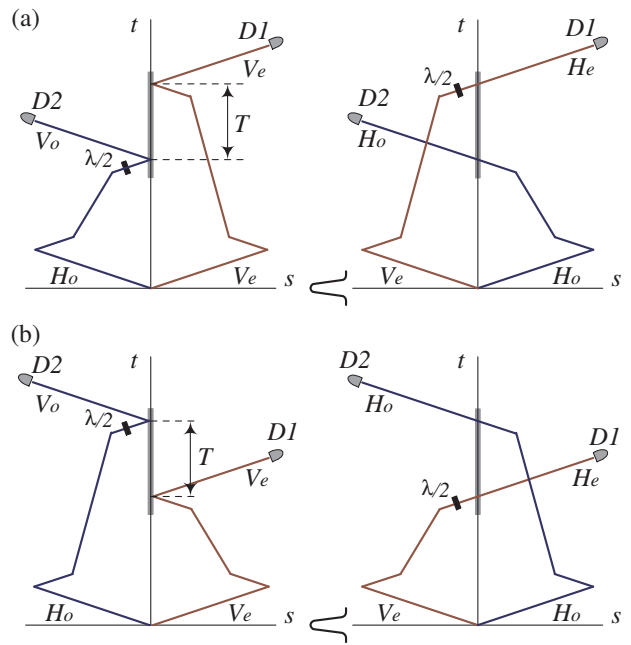}
\caption{\label{fig:feynman}Possible quantum mechanical amplitudes for a photon
pair can take when the optic axes of (a) both QR1 and QR2 are oriented
vertically, (b) both QR1 and QR2 are oriented horizontally. Vertical
(horizontal) axis represents time (space). Thick gray line represents the
polarizing beamsplitter (PBS).}
\end{figure}

Let us first consider the case in which SA/HOM effect is observed even though
the photons never overlap at the beamsplitter (the arrival time difference
between the photons at the beamsplitter is much greater than the coherence
times of the pump photon and the SPDC photons). This case can be realized by
setting the optic axes of both QR1 and QR2 vertically. As explained before,
there are two possibilities for the polarization state photon pair;
$|H_o\rangle|V_e\rangle$ or $|V_e\rangle|H_o\rangle$. $|H\rangle$ and
$|V\rangle$ refer to the orientation of the polarization of the photon,
horizontal and vertical, respectively and the subscripts $e$ and $o$ refer to
whether the photon belongs to the e-ray or o-ray of the crystal, initially. For
example, $|H_o\rangle$ refers to the photon polarized horizontally and belongs
to the o-ray of the crystal. Note, however, that $|H_e\rangle$ can never occur
due to the orientation of the BBO crystal.

Since the optic axes of both quartz rods are oriented vertically (i.e., fast
axis oriented horizontally), a horizontally polarized photon experiences
relatively less group delay with respect to the vertically polarized its twin.
This relative delay is calculated to be approximately $T\approx 630$ fsec for
20 mm long quartz rods used in this experiments. This delay $T$ is much bigger
than 130 fsec pump pulse coherence time and the coherence times of the SPDC
photons which are defined by the bandwidth of the interference filters:
$\tau\sim\lambda^2/(c \cdot \Delta \lambda)\approx 100$ fsec. Note that the
delay $T$ is different from the relative delay between the two arms of the
interferometer which is introduced by moving P1.

\begin{figure}[tbp]
\includegraphics[width=3.1in]{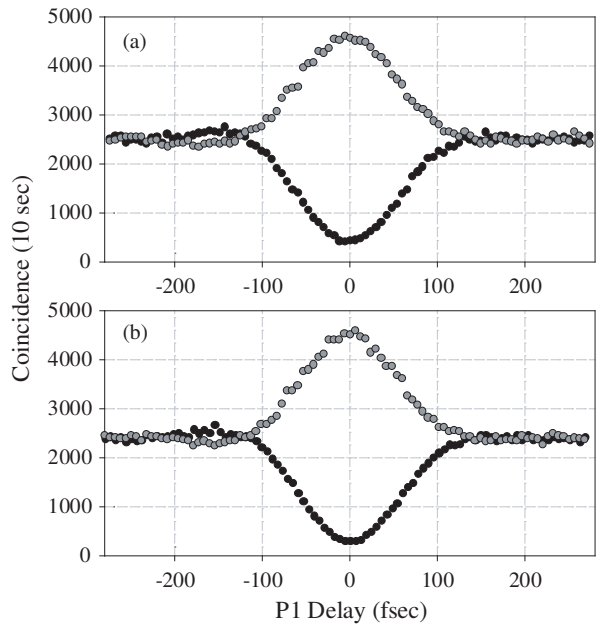}
\caption{\label{fig:data}High visibility quantum interference is observed (a)
when both QR1 and QR2 are oriented vertically ($V\sim83\%$), see
Fig.~\ref{fig:feynman}(a) and (b) when both QR1 and QR2 are oriented
horizontally ($V\sim87\%$), see Fig.~\ref{fig:feynman}(b). `Dip' (dark circle)
is observed for $45^\circ/45^\circ$ and `peak' (gray circle) is observed for
$45^\circ/-45^\circ$ analyzer angles (A1/A2).}
\end{figure}

This situation is well represented in the Feynman-like space-time diagram shown
in Fig.~\ref{fig:feynman}(a). Since the HWP transforms the polarization state
$|H\rangle \leftrightarrow |V\rangle$, there are only two possible two-photon
amplitudes: both photons reflected (r-r) or both photons transmitted (t-t). It
is not hard to see that the arrival time difference between the photon pair at
the beamsplitter, $T$, is much bigger than both the coherence times of the
photons themselves, $\tau$, and the pump pulse. However, if the both arms of
the interferometer have the same length (P1 delay = 0 fsec), the amplitudes r-r
and t-t cannot be distinguished by the arrival times of the photons (even with
infinitely fast photodetectors). The only distinguishing information for the
two amplitudes is in their polarization and it can be erased by setting the
polarization analyzers either at A1/A2 = $45^\circ/45^\circ$ or at
$45^\circ/-45^\circ$ \cite{eraser}. Therefore, even though the two photons
never overlap at the beamsplitter and the arrival time difference is much
bigger than the pump coherence time, SA/HOM effect may still occur. This is
because the SA/HOM effect, in general, is the result of indistinguishability
between two two-photon amplitudes but not due to `photon bunching at
beamsplitter' effect.

We can also consider when both QR1 and QR2 are horizontally oriented. In this
case, nothing is changed except that the delays experienced by each photons are
reversed. The two-photon amplitudes for this case can be seen in
Fig.~\ref{fig:feynman}(b). It is clear that the two-photon amplitudes remain
indistinguishable however the order in which the detectors fire has reversed.
In Fig.~\ref{fig:feynman}(a), D2 always fires before D1 by time $T$. In
Fig.~\ref{fig:feynman}(b), D1 always fires before D2 by the same amount of
time.

The experimental data for these two cases are shown in Fig.~\ref{fig:data}.
When taking the data, we fixed the orientations of the quartz rods and scanned
the interferometer arm delay by moving the trombone prism P1. This procedure
was repeated for different orientations of quartz rods for two different
analyzer settings: A1/A2 = $45^\circ/45^\circ$ and $45^\circ/-45^\circ$. The
observed visibilities are higher than the classical limit (50\%) as well as the
limit for the Bell-inequality violation (71\%) which clearly establishes that
the observed interference is of quantum origin.

\begin{figure}[tbp]
\includegraphics[width=3.1in]{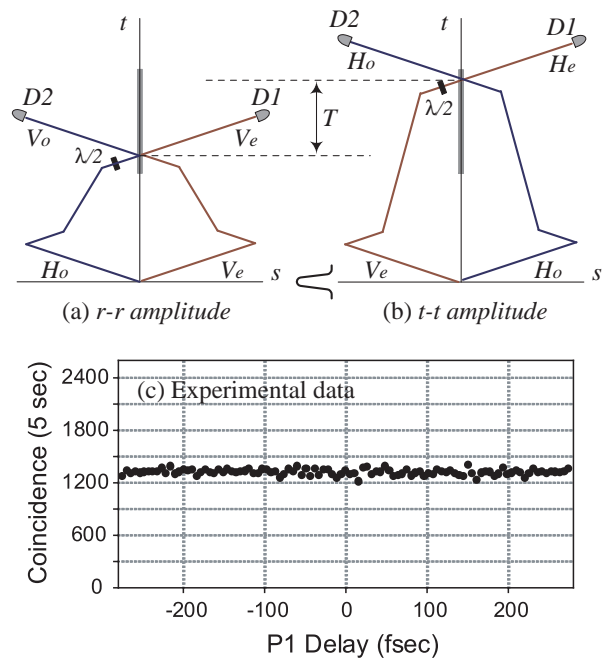}
\caption{\label{fig:feynman2}Possible quantum mechanical amplitudes when the
optic axis for QR1 (QR2) is set at vertical (horizontal). Individual photons do
overlap at the beamsplitter in both amplitudes. However, due to the intrinsic
distinguishability between the two amplitudes, quantum interference (SA/HOM
effect) cannot occur. (c) Experimental data showing no interference. Analyzer
angles are $45^\circ/45^\circ$. }
\end{figure}

Let us now consider the case in which two down-converted SPDC photons do
overlap at the beamsplitter, yet no quantum interference (SA/HOM effect) can
occur. To consider this case, we need to choose orientations of the quartz rods
other than both vertical and horizontal. Here we consider QR1 = V and QR2 = H.
In this case, the photon pair experiences the same group delay in both arms of
the interferometer because the photon pair has the polarization state
$|H\rangle|V\rangle$ or $|V\rangle|H\rangle$. The Feynman diagram for this case
can be seen in Fig.~\ref{fig:feynman2}(a) and Fig.~\ref{fig:feynman2}(b). It is
clear that the individual photons do overlap at the beamsplitter for both r-r
and t-t amplitudes. However, the two amplitudes are intrinsically
distinguishable because if we had infinitely fast detectors, the pump pulse
would act as a clock and we would then be able to distinguish the two
amplitudes. However, should SA/HOM effect be a result of `photon bunching', a
dip or peak in coincidence counts should occur. We have done this experiment
and observed no interference for any polarizer settings, see
Fig.~\ref{fig:feynman2}(c). This clearly shows that the photon bunching picture
often used in literature is indeed incorrect in general and should not be used
whenever possible.

Note that it is, however, possible to observe interference if the pump pulse
coherence time is bigger than $T$ in Fig.~\ref{fig:feynman2}. The uncertainty
provided by a long coherence time of the pump pulse would then make the two
amplitudes indistinguishable, thus leading to interference. We are then back to
the situation where the photon bunching picture and the quantum amplitude
picture are both valid. This situation is then similar to Pittman et al's
scheme. It is therefore necessary that all relevant coherence times should be
much smaller than the photon arrival time difference at the beamsplitter to be
able to make a clear distinction between the two pictures.

To summarize, we reported a quantum interference experiment in which
two-photon quantum interference was observed even though the photon pair
arrival time difference at the beamsplitter was much bigger than the coherence
times of the individual photons as well as the pump pulse. We have also
discussed the case in which photons did overlap at the beamsplitter but no
quantum interference could be (had been) observed. This experiment clearly
demonstrates that SA/HOM effect is indeed due to indistinguishability of
two-photon amplitudes but not due to the `photon bunching' effect of individual
photon wavepackets. It also demonstrates that genuine higher-order interference
effects should not and cannot be explained by using lower-order interference
picture.

Dirac, in his famous textbook, stated ``Each photon then interferes only with
itself \cite{dirac}." In two-photon interference experiments, we may then say
``Two-photon or biphoton interferes only with itself \cite{nonlocal0}."

Finally, we note that this work may be of some use in quantum cryptography and
in studying decoherence management in entangled two-qubit systems as we observe
near complete restoration of quantum interference (without any post-selection
in principle) after the qubit pairs (which are in mixed states), generated by a
femtosecond laser pulse, went through certain birefringent elements
\cite{newsource}.

The author wishes to thank C. D'Helon, W.P. Grice, and Y. Shih for helpful
comments. This research was supported in part by the U.S. Department of Energy, Office
of Basic Energy Sciences, the National Security Agency, and the LDRD program of the Oak Ridge
National Laboratory, managed for the U.S. DOE by UT-Battelle, LLC, under
contract No.~DE-AC05-00OR22725.


\begin{thebibliography}{}
\bibitem{note} We note that, unlike common misconception, the photon pairs found
in these two directions are not polarization entangled. In fact, the
polarization state of the photon pair is in a mixed state \cite{notentangled}.
\end{thebibliography}
\end{document}